\documentclass[conference]{IEEEtran}
\IEEEoverridecommandlockouts

\RequirePackage{etex}
\usepackage[pdftex]{graphicx}
\graphicspath{{../pdf/}{../jpeg/}}
\DeclareGraphicsExtensions{.pdf,.jpeg,.png}

\usepackage[cmex10]{amsmath}
\usepackage{algorithmic}
\usepackage{array}
\usepackage{mdwmath}
\usepackage{mdwtab}
\usepackage{eqparbox}
\usepackage{url}
\usepackage{tikz}
\usetikzlibrary{matrix,chains,positioning,decorations.pathreplacing,arrows}
\usepackage{mathtools}
\usepackage{tikz-network}
\usepackage{booktabs}

\usepackage[hidelinks]{hyperref}
\hypersetup{
    colorlinks=false,
    linkcolor=blue,
    filecolor=magenta,      
    urlcolor=cyan,
}

\ifCLASSOPTIONcompsoc
    \usepackage[caption=false,font=footnotesize,labelfont=sf,textfont=sf]{subfig}
\else
    \usepackage[caption=false,font=footnotesize]{subfig}
\fi

\usepackage{cite}
\usepackage{amsmath,amssymb,amsfonts}
\usepackage{algorithmic}
\usepackage{graphicx}
\usepackage{textcomp}
\usepackage{xcolor}
\def\BibTeX{{\rm B\kern-.05em{\sc i\kern-.025em b}\kern-.08em
    T\kern-.1667em\lower.7ex\hbox{E}\kern-.125emX}}

\begin{document}

\onecolumn

\noindent \textcopyright{} 2019 IEEE. Personal use of this material is permitted. Permission from IEEE must be obtained for all
other uses, in any current or future media, including reprinting/republishing this material for advertising or
promotional purposes, creating new collective works, for resale or redistribution to servers or lists, or reuse
of any copyrighted component of this work in other works.

\twocolumn

\title{The Validation of Graph Model-Based, Gate Level Low-Dimensional Feature Data for Machine Learning Applications \\
\thanks{This work was supported by the RESCUE ETN project. The RESCUE ETN project has received funding from the European Union's Horizon 2020 Programme under the Marie Skłodowska-Curie actions for research, technological development and demonstration, under grant No. 722325}
}

\author{%
\IEEEauthorblockN{%
  Aneesh Balakrishnan\IEEEauthorrefmark{1}\IEEEauthorrefmark{2},
  Thomas Lange\IEEEauthorrefmark{1}\IEEEauthorrefmark{3},
  Maximilien Glorieux\IEEEauthorrefmark{1},
  Dan Alexandrescu\IEEEauthorrefmark{1},
  Maksim Jenihhin\IEEEauthorrefmark{2}%
}
\IEEEauthorblockA{%
  \IEEEauthorrefmark{1}\textit{iRoC Technologies}, Grenoble, France \\
   \IEEEauthorrefmark{2}\textit{Department of Computer Systems, Tallinn University of Technology}, Tallinn, Estonia \\
  \IEEEauthorrefmark{3}\textit{Dipartimento di Informatica e Automatica, Politecnico di Torino}, Torino, Italy \\
  \{aneesh.balakrishnan, thomas.lange, maximilien.glorieux, dan.alexandrescu\}@iroctech.com \qquad
  maksim.jenihhin@taltech.ee}
}

\IEEEoverridecommandlockouts
\IEEEpubid{\makebox[\columnwidth]{978-1-7281-2769-9/19/\$31.00~\copyright2019 IEEE \hfill} \hspace{\columnsep}\makebox[\columnwidth]{ }}

\maketitle

\begin{abstract}
As an alternative to traditional fault injection-based methodologies and to explore the applicability of modern machine learning algorithms in the field of reliability engineering, this paper proposes a systemic framework that explores gate-level netlist circuit abstractions to extract and exploit relevant feature representations in a low-dimensional vector space. A scalable feature learning method on a graphical domain called node2vec algorithm \cite{node2vec} had been utilized for efficiently extracting structural features of the netlist, providing a valuable database to exercise a selection of machine learning (ML) or deep learning (DL) algorithms aiming at predicting fault propagation metrics. The current work proposes to model the gate-level netlist as a Probabilistic Bayesian Graph (PGB) in the form of a Graph Modeling Language (GML) format. To accomplish this goal, a Verilog Procedural Interface (VPI) library linked to standard simulation tools has been built to map gate-level netlist into the graph model. The extracted features have used for predicting the Functional Derating (FDR) factors of individual flip-flops of a given circuit through Support Vector Machine (SVM) and Deep Neural Network (DNN) algorithms. The results of the approach have been compared against data obtained through first-principles approaches. The whole experiment implemented on the features extracted from the 10-Gigabit Ethernet MAC IEEE 802.3 standard circuit.

\end{abstract}

\begin{IEEEkeywords}
Probabilistic Graph Model, Deep learning, Machine Learning, Functional Derating, Single-Event Upset (SEU), Gate-Level Netlist, Graph Modeling Language.
\end{IEEEkeywords}

\section{Introduction}

System engineering focuses on the integration of new small-scale technologies, constantly advancing the state of the art. The costly and difficult implementation of micro- and nano-scale devices highlights the challenges faced by all the partners from the design and manufacturing flow and, always aiming at improving their technological competitiveness. Current quality requirements, from end users or industrial standards, motivate designers and reliability engineers to dedicate significant effort and resources to reliability and functional safety aspects. Particularly, issues due to radiation based effects - Single Event Effects (SEEs) impact reliability metrics and are challenging to evaluate. A valuable approach to tackle these effects is the fault injection and simulation principle that provides precise and accurate information about circuit behaviour under stress, allowing the calculation of actual circuit-level reliability metrics.

\subsection{Motivation}

Nowadays, increased user expectations or actual factual requirements formulated by the reliability and functional safety standards in high dependability applications make reliability modeling and assessment increasingly relevant. The reliability assessment process is usually accomplished with the different types of fault injection methods like exhaustive and random. The exhaustive fault injection method is obviously the ultimate reliability assessment method in terms of accuracy but very cumbersome in terms of resources, time, EDA licenses and so on, making this approach unfeasible on medium and large circuits. The random fault injection provides a solution to avoid unreasonable costs while allowing for accuracy (or statistical significance) on the proposed scope. Research proposals based on mathematical and statistical methods are always put forward by the research scientists. Nowadays, ML/DL techniques \cite{William},\cite{Michal},\cite{xgboost}, \cite{breiman2001random}, \cite{Franco} are more advanced and greatly favoured by researchers to learn statistical and functional dependencies between the feature representations of different systems. This is the main motivation for the idea of getting different algorithms and trying to find the best ways to develop relevant feature databases in the field of reliability assessment.

\subsection{Organization of the Paper}

The paper includes five sections in total. Section I summarized the State-Of-The-Art in the field of reliability engineering before presenting the motivation of the current work and the organizational structure of the paper. Section II gives a background introduction to Support Vector Machine, Deep Neural Networks and, different reliability factors of the microelectronic systems. In Section III, the main methodological implementation overview has given. Also, it explains the node2vec algorithm and different regression metrics. Section IV illustrates the results and their validations in terms of different regression metrics and diagrams. As future progress of this work, a deep learning algorithm with a complex architecture called GCN has introduced and, its recent progress has briefed in section V. In VI, the whole work and its holistic approaches have concluded.


\section{Background}
\subsection{Interpretation of Standard Reliability Based Terms}
\subsubsection{\textbf{Single Event Effects}} 
As the term suggests, a single event effect \cite{Soft:Errors} results from a Single Event - the interaction of a energetic particle with the device. The main effects are classified in two categories, destructive and non-destructive. This work is mainly contributing to the derating analysis of Single Event Upsets in sequential elements. The quantitative analysis of single event effects is based on different derating factors, called functional derating, logical derating, temporal derating, and electrical derating. 

\subsubsection{\textbf{Electrical Derating}}
    The Electrical Derating evaluates the propagational probability of the analog Single Event Transient (SET) pulse generated by the particle interaction. Based on their electrical pulse width and electrical amplitude range, it defined how well a transient error obstructs the standard signal propagation in the given circuit.   
    
\subsubsection{\textbf{Temporal Derating}}
    Temporal (or time) derating represents the opportunity window of an event (SET or SEU) and it's probability to be latched to the downstream sequential elements like flip-flop, latch and memory.   

\subsubsection{\textbf{Logical Derating}}
    The logical vulnerability of the SEE within the combinational (or) sequential cell networks based on their logical boolean functions is quantified with masking effect probability, termed as logical derating factor.  

\subsubsection{\textbf{Soft Error}}
    The fault - the primary consequence of the Single Event (SET/SEU) can be dropped or blocked in the circuit. If the fault propagates to and is memorized in state element (flip-flop, latch, memory), then it becomes a Soft Error. Please note that Bit/Cell Upsets (Single or Multiple) in memory instances are also Soft Errors.
    
\subsubsection{\textbf{Functional Derating}}
    Functional Derating evaluates how likely is the Soft Error to cause an observable impact (Functional Failure) on the functioning of the circuits or systems. 

\subsection{The Machine learning and Deep Learning Algorithms}

Giving an introductive subsection helps the reader to develop a clear idea of the relationship between Artificial Intelligence (AI), Machine Learning and Deep Learning. Deep Learning or Deep Neural Networks (DNN) or Artificial Neural Networks (ANN) are commonly considered as a subset of machine learning which in turn is derived from the concept of Artificial Intelligence. In machine learning, a database including the labeling vector of each class is parsed during the learning process and then exploit the learned dependencies between feature and class labels for deriving a decision margin, whereas, in case of deep learning algorithms, it appears in layers that can learn and make intelligent decisions on its own. 

\subsubsection{\textbf{Support Vector Machine}}

The support vector machine works on the foundation of a good theoretical learning algorithm to solve regression analysis as well as classification type problems. The SVM for regression analysis can be called as SVR in short.  It was invented by Vladimir Vapnik and his co-workers, and first introduced at the Computational Learning Theory (COLT) 1992 conference with the paper \cite{Vapnik}. SVM characterizes the maximal margin algorithm for supervised learning models. In the maximal margin principle, SVR tries to find the optimal hyperplane which maximizes the margin and minimizes the error. Compared to classification problems, regression analysis outputs a continuous variable. The SVR approach defines a margin of tolerance $\epsilon$ where no penalty is given to errors. At the same time, it punishes the wrong estimation with a cost-insensitive symmetric hinge loss function.   

An alternate approach in SVR applications is the kernel modification. A kernel which possible to transform the given data set to higher dimensional space to derive a liner decision boundary. A properly chosen Radial Basis Functions (RBF) had employed as a kernel function in this work. RBF is also called the Gaussian Kernel which means that each feature vector of the dataset in the transformed dimensional space influenced by the Gaussian observation. 

\subsubsection{\textbf{Deep Neural Network}}

Deep Neural Network is an important step in the machine learning algorithms. Their learning methods are trying to model data with complex architectures and distributions by combining different non-linear transformations. In this work, a general fully connected DNN is implemented. The other main categories of deep learning methods are Convolutional Neural Network (CNN) and Recurrent Neural Networks (RNN). The elementary bricks of deep learning are the artificial neurons (perceptrons) which are inspired by biological neurons.  An artificial neuron combines the input signals with adaptive weights and uses an activation function to deliver the output to be estimated. An in-depth discussion about the architecture as well as the adopted parameters has given in section IV.

\section{Methodology}
\subsection{Overview of the Work}

\begin{figure}[ht]
    \includegraphics[width=\linewidth]{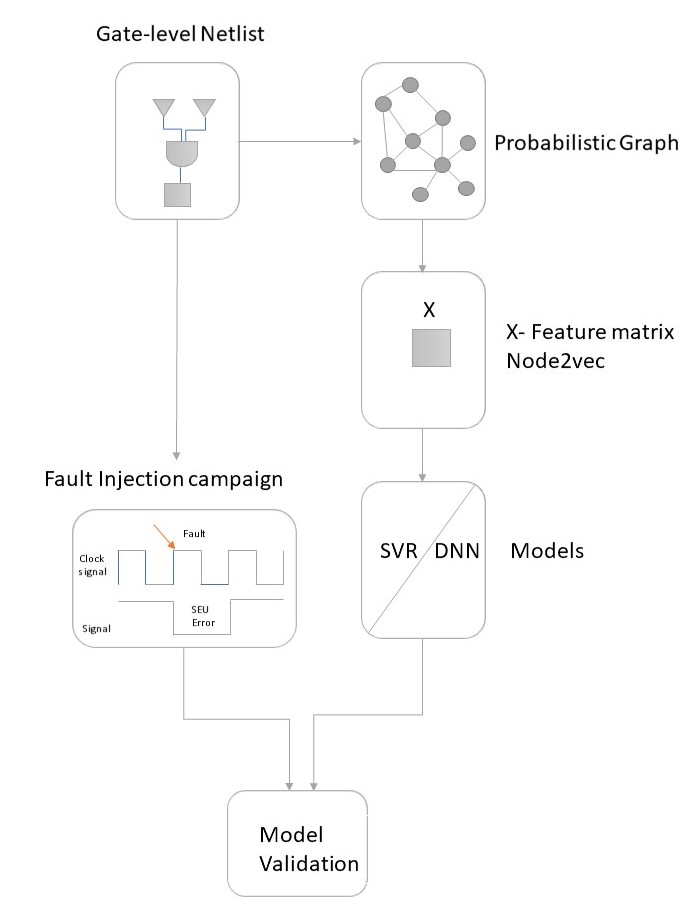}
    \caption{Systematic block diagram of the scientific work }
    \label{figure_block}
\end{figure}

A global overview of the work is portrayed in figure \ref{figure_block}. We start by mapping gate-level netlist into the probabilistic graph model. As the structural information of gate-level netlist is transformed into the probabilistic graph, the statistical properties of a graph node conventionally equivalent to that of a sequential (flip-flop) or logic (gate) element of the circuit are exposed. To execute this preliminary part of the work, different user-defined  VPI  functions had been written in C/C++ and linked to standard EDA logic simulators. The VPI library is able to extract all the relevant details of the gate-level netlist and formats them into a probabilistic graph model through GML graph attributes. In the next stage of the work, an SVM-Regressor (SVR) - a standard machine learning algorithm and fully connected DNN based on the deep learning algorithm, were adopted as the learning-frameworks of the features from the probabilistic graph.\\
The feature matrix X for the implemented learning-frameworks is obtained by the random walk method using the node2vec algorithm. This algorithm can provide the feature dataset for the Circuit Under Test in a desired dimensional space within fractions of seconds. The random walk method gives a feature vector corresponding to a node by preserving neighborhood structure. The feature vector is mainly based on transition probabilities from source to target nodes in the neighborhood area and also the degree of nodes.

We choose a first principle approach - a basic, straightforward fault injection and simulation campaign, as a reference model and as a comparison baseline. This way, more stringent validation of the expected goals became possible. As observed from figure  \ref{figure_block}, the fault injection based ground truth data is shuffled and has split with a test size of 40\% and a Training size of 60\%. After training the learning models, predicted FDR values of flip-flops has been compared with the test vectors from the fault injection campaign FDR data. The ML/DL algorithms had been implemented in python with the help of Keras and Scikit-learn libraries which are available as open-source machine learning libraries for the Python programming language.

\subsection{Node2vec: Scalable Feature Learning on Graphs}

The node2vec algorithm proposed by Aditya Grover in \cite{node2vec} is endowed here in its novelty.  The node2vec algorithm is a framework for learning continuous feature representations in the graph network. It maps the nodes in the graph into the desired dimensional feature space which maximizes the likelihood of preserving the network neighbourhood of nodes. Node2vec algorithm can apply to any given directed or undirected, weighted or unweighted edge networks.     

Nowadays, representing a dataset in a graphical domain becomes a very useful (and obligatory) tool. We use this approach for predicting and visualizing the probability factors over nodes and edges. The netlist from the gate-level abstraction of the circuits is successfully represented in the graph domain. For performing a prediction analysis, a careful effort is required to develop a feature vector space that suitable for different learning algorithms. This requirement has achieved with the node2vec algorithm.

The feature learning framework of the node2vec algorithm had been formulated as a maximum likelihood optimization problem. The given network can be represented as $G = (\nu,\varepsilon)$, where $\nu$ represents vertices or nodes and $\varepsilon$ represents the edges between the vertices. $f : V \rightarrow \mathbb{R}^{d}$ is the mapping function from a node to $d$ dimensional feature space, where $V$ stands for a whole set of vertices. $f$ is a matrix with size of $|V| \times d$. A neighborhood sampling strategy $S$ is used to define a network neighbourhood as $ N_{s}(u)$ of a source node $u$. The framework optimizes the objective function $f$ by maximizing the log-probability of observing a network neighbourhood $N_{s}(u)$ for a node $u$, conditioned on its feature representation. The objective function is given by:

\begin{equation}
\label{layer}
     \displaystyle \max_{f} \displaystyle \sum_{u \epsilon V}{log Pr\left( N_{s}(u) | f(u)\right)}.
\end{equation}

The sampling strategy developed for node2vec is a flexible random walk that interpolates two important sampling strategies termed as Breadth-first Sampling (DFS) and Depth-First Sampling (DFS). In BFS, the sampling nodes are the very immediate neighbors of the source node whereas, in DFS the neighbors have been obtained by sampling sequentially at increasing distance from a source node. The two important factors in the node2vec algorithm are flexible biased random walk and search bias $\alpha$. Let consider a source node $u$ and a random walk length $l$ and $c_i$ denote the $i^{th}$ node in the walk from source node $c_0 = u$. The probability of $c_i$ given $c_{i-1}$ is generated by:  
 
\begin{equation}
\label{layer-1}
P(c_i =x\ |\ c_{i-1}=v ) = \left\{
       \begin{array}{cc}
           \frac{\pi_{vx}}{Z} & \textrm{if} \quad (v,x) \in E  \\
           0 & \textrm{Otherwise}
       \end{array} 
       \right\}
\end{equation}

where, where $\pi_{vx}$ is the unnormalized transition probability between nodes v and x, and Z is the normalizing constant.

\begin{figure}[ht]
    \includegraphics[width=\linewidth]{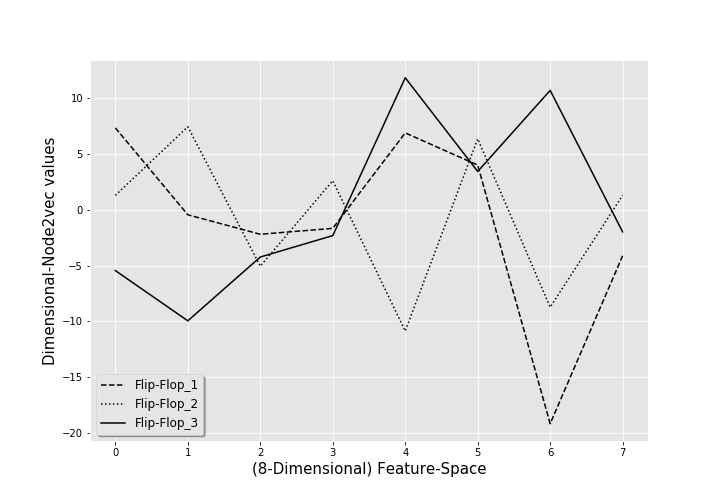}
    \caption{Feature vector of three arbitrary flip-flops}
    \label{fig:dim}
\end{figure}

The search bias factor $\alpha$ is a major factor in calculating $\pi_{vx}$.  Consider a random walk that just traversed the edge (t,v) and resides on node v. As a next step in the random walk, an unnormalized transition probability $\pi_{vx}$ on the edge (v,x) leading from v, is estimating. The unnormalized transition probability is set to $\pi_{vx} = \alpha_{pq}(t,x).w_{vx}$, where: 

\begin{equation}
\label{layer-3}
\alpha_{pq}(t,x) = \left\{
       \begin{array}{cc}
           \frac{1}{p} & \textrm{if}\ d_{tx} = 0 \\
           1 & \textrm{if}\ d_{tx} = 1\\
           \frac{1}{q} & \textrm{if}\ d_{tx} = 2 \\
       \end{array} 
       \right\}
\end{equation}
and $w_{vx}$ is the weight of the edge. In the case of unweighted edge, $w_{vx} = 1$. The $d_{tx}$ is the shortest path between t and x. Parameter $p$ is called the Return Parameter and it controls the likelihood of immediately revisiting node in the walk. $q$ is called an In-Out parameter which allows the search to differentiate between inward and outward nodes. Here, feature space with dimension 8 had extracted. The feature vectors of three arbitrary flip-flops had plotted in figure \ref{fig:dim} for giving an illustration of the vector's statistical variance.

\subsection{Regression Evaluation Metrics}
\subsubsection{\textbf{Mean Squared Error (MSE)}}

If $\hat{y}_i$ is the predicted value and $y_i$ is the true value corresponding to the $i^{th}$ sample, then the mean squared error to be estimated over $n$ samples is defined as,

\begin{equation}
\label{mse}
    MSE(y,\hat{y}) = \frac{1}{n} \displaystyle \sum_{i=0}^{n-1}(y_i - \hat{y_i})^2
\end{equation}

The regression error will become minimal as MSE approaches to zero.
\subsubsection{ \textbf{R - Squared Score ($R^2$)}} also known as the coefficient of determination. If $\hat{y}_i$ is the predicted value of the $i^{th}$ sample, and $y_i$ is the corresponding true value, then the coefficient of determination estimated over $n$ samples defined as,

\begin{equation}
\label{R}
    R^2(y,\hat{y}) = 1- \frac{\displaystyle \sum_{i=0}^{n-1}(y_i - \hat{y_i})^2}{\displaystyle \sum_{i=0}^{n-1}(y_i - \Bar{y_i})^2}  \\
\end{equation}

where, $\Bar{{y_i}} = \frac{1}{n} \displaystyle \sum_{i=0}^{n-1} y_i $. Numerical value 1 indicates a good regression fit, while 0 indicates a worse fit.

\subsubsection{\textbf{Explained variance score (EVS)}}

f $\hat{y}$ is the predicted value of the target value $y$, then Explained variance score estimated over $n$ samples is defined as,

\begin{equation}
\label{evs}
    EVS(y,\hat{y}) = 1- \frac{Var\{y-\hat{y}\}}{Var\{y\}}
\end{equation}

where, $Var$ is the square of the standard deviation. The best possible score is 1.


\section{Result : Modeling and Validations}
To test the applicability of the node2vec based features for the machine learning frameworks in the system reliability evaluation, a validation effort has been performed on a 10-Gigabit Ethernet MAC IEEE 802.3 standard circuit. Experimenting with fault injection for each flip-flop independently and documenting how probable is the fault to affect the overall function of the circuit as the Functional Derating factor provides the reference dataset for the validation. About one thousand two hundred and two (1202) flip-flops have used for evaluating the prediction models. The circuit is accessible at OpenCores as the 10-Gigabit Ethernet project.  
 
\begin{figure}[ht!]
\centering
\subfloat[Prediction over 40\% Test data]
{
    \includegraphics[clip,trim=50 25 0 0,scale=0.26]{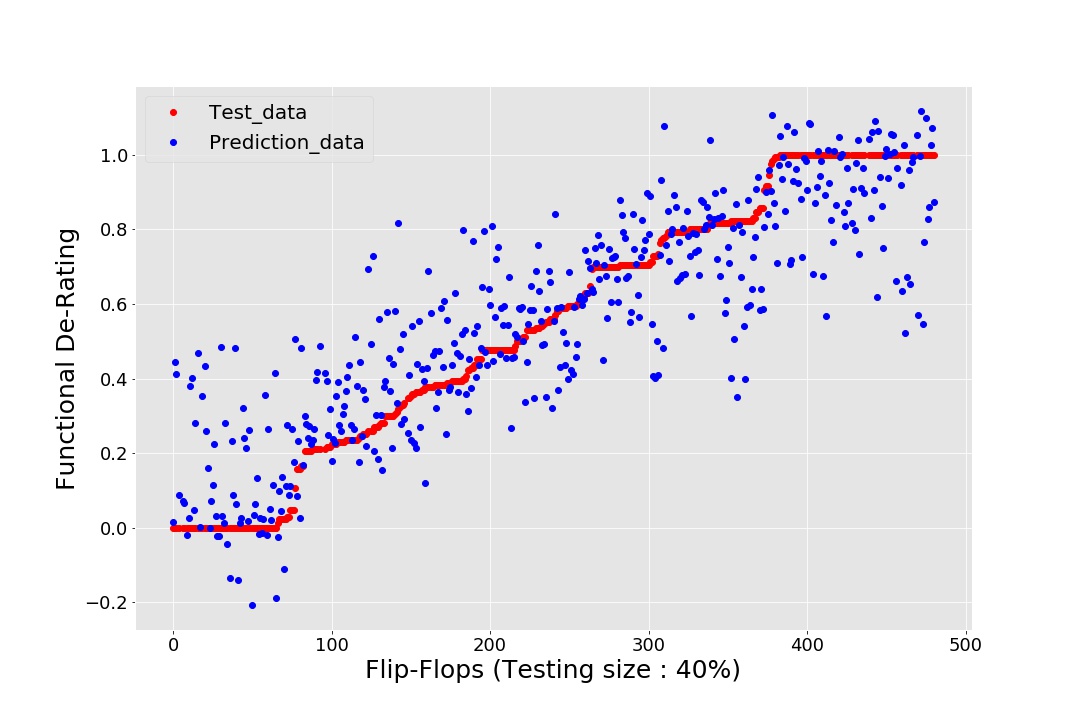}
    \label{fig:svm_pre}
}\\
\subfloat[Scatter plot between prediction and true value]
{
    \includegraphics[clip,trim=50 25 0 0,scale=0.26]{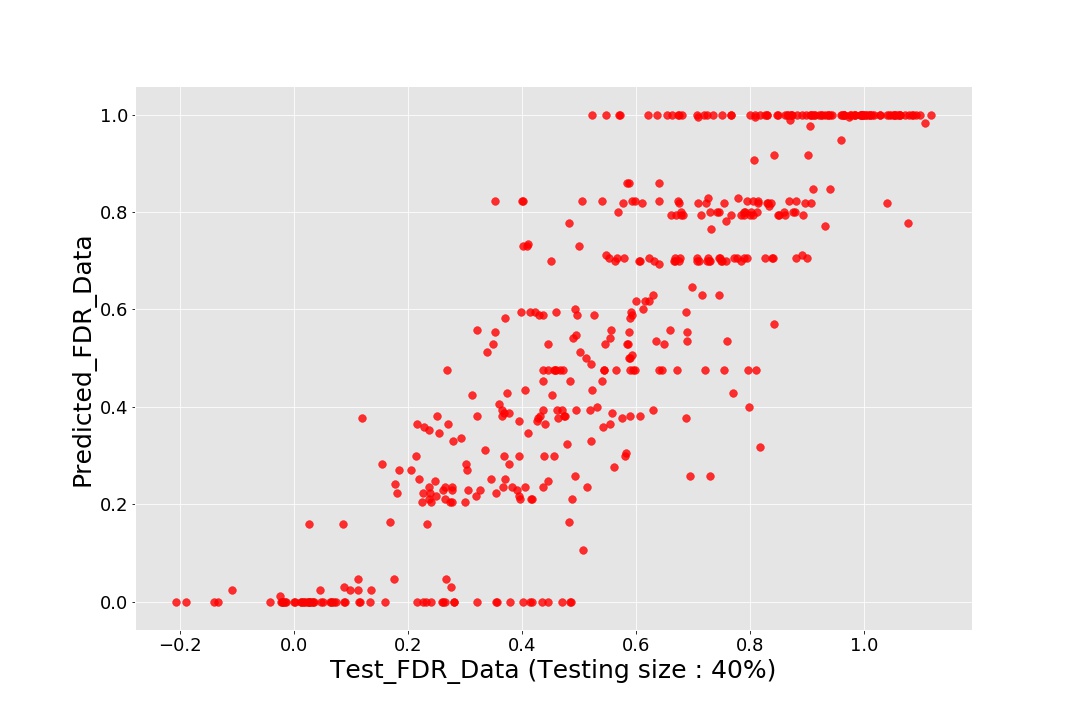}
    \label{fig:svm_scatter}
}
\caption{Regression by SVR model}
\label{fig:svm_pre_scatter}
\end{figure}

\begin{figure}[ht!]
\centering
\subfloat[Prediction over 40\% Test data]
{
    \includegraphics[clip,trim=50 25 0 0,scale=0.26]{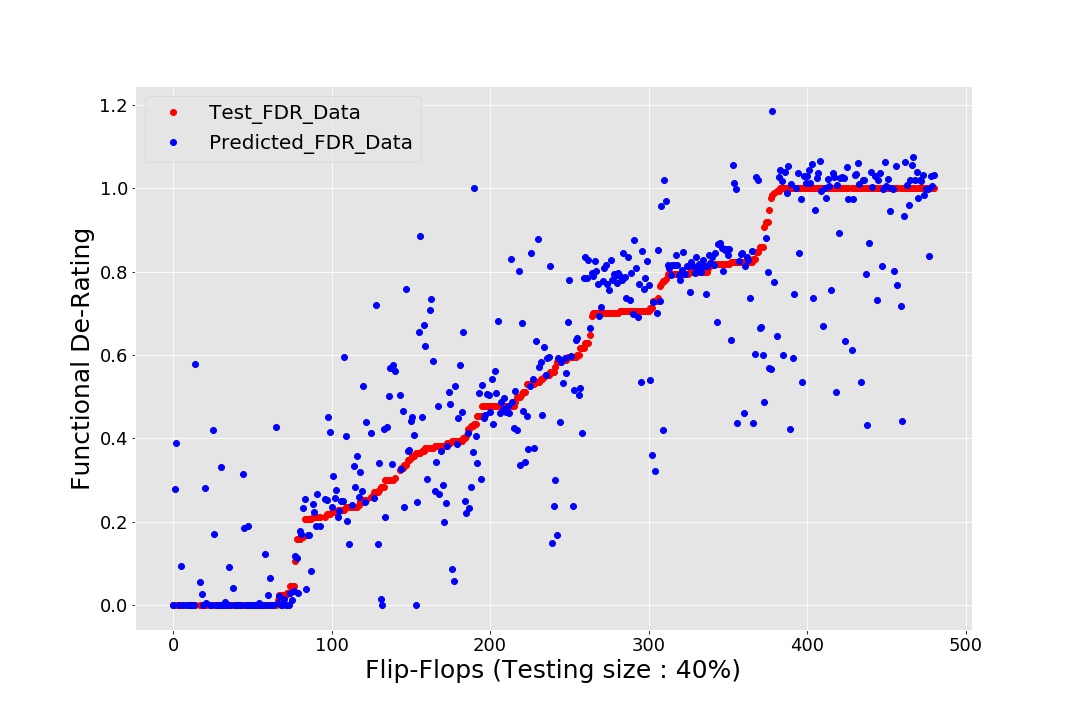}
    \label{fig:dnn_prediction}
}\\
\subfloat[ Scatter plot between prediction and true value]
{
 \includegraphics[clip,trim=50 25 0 0,scale=0.26]{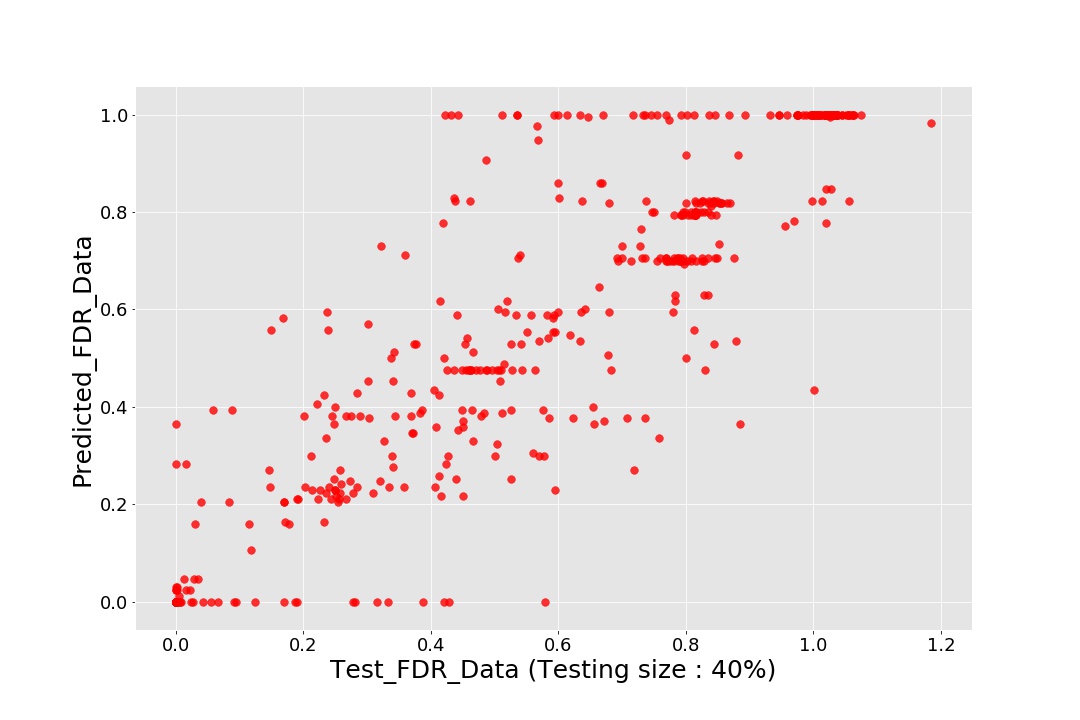}
 \label{fig:dnn_scatter}
}

\caption{ Regression by DNN Model}
\label{fig:dnn_pre_scatter}
\end{figure}

\subsection{Result Analysis : SVR}

The prediction result of Support Vector Machine Regression are provided in figure \ref{fig:svm_pre} and jointly presented a scatter plot in figure \ref{fig:svm_scatter} respectively. The corresponding evaluation metrics have been given in Table \ref{tab1}. In SVR, we use the RBF kernel function which described as,  

\begin{equation}
\label{}
K(X,X') = \exp \left({-\gamma \left\Vert X - X' \right\Vert }^2 \right).
\end{equation}

The $X\ \textrm{and}\ X'$ are the two data points in vector form. The kernel K maps them to higher dimensional vector space. $\gamma$ is called the spread of the kernel function and, it tuned to $\gamma$ = 0.01. The other important parameter is epsilon $\epsilon$ which, responsible for error tolerance and set to $\epsilon$ = 0.0125. The parameter $C$ is the regularization scheme and, proper value is chosen for the penalty factor $C$. Here $C$ = 10. A grid-search cross-validation method tunes the parameter values. From the prediction diagram \ref{fig:svm_pre}, the predicted values approximating the original values which, sorted in ascending order by values. The scatter plot in figure \ref{fig:svm_scatter} indicating a good correlation between predicted and original test data. But there is still a space for improvement because the scatter plot having a variance between the axial components. The metrics $R^2$ from Table \ref{tab1} is indicating the good regression fit of prediction with original data. It is almost 69\%. If the predicted values approximate more likely to the tested data, the $R^2$ will tend to the numerical value 1. In the same way, the metric MSE form Table \ref{tab1} is equal to 0.027 and, it will close to 0 when the approximation becomes better. The metric EVS also mentioned in Table \ref{tab1}.

\subsection{Result Analysis : DNN }

The DNN architecture had been chosen according to table \ref{tab3}. The input layer is nothing but the feature vectors. The Dense\_1, Dense\_2, Dense\_3, and Dense\_4 are the hidden layers. Dense\_5 is called the output layer which outputs the estimated regression values. Each hidden layer is a fully-connected dense layer where the number of inputs to each neuron is equal to the output size of the previous layer. The weights of neuron inputs and the bias factor are the parameters that need to be optimized. The hyper-parameters termed as loss = 'Mean Squared Error', optimizer = 'Adam' \cite{Kingma:2014vow} and batch\_size = 10 are chosen according to cross-validation method. The Dense\_1 layer has a shape of 126 neurons. With the input feature vector of dimension 8, DNN training for getting a good prediction accuracy becomes difficult. Therefore, the Dense\_1 layer will map the low dimensional input vectors to a high dimensional space. DNN will show significant performance with a higher dataset dimension.

\begin{table}[htbp]
    \centering
    \caption{DNN Architecture}
    \label{tab3}
    
    \begin{tabular}{lccccccc}
    \toprule
        Layer &  Output shape &
            Parameters \\
    \midrule
        Dense\_1 &  126 &
            25326  \\
        Dense\_2 & 64 &
            8128  \\
        Dense\_3 & 36 &
            2340  \\
        Dense\_4 & 12 &
            444  \\
        Dense\_5 & 1 &
            13  \\
    \bottomrule
    \end{tabular}
\end{table}

Figure \ref{fig:dnn_prediction} shows that the DNN method provides an adequate prediction. A majority of the estimations are close to the true values, which indicates a good $R^2$ value. From Table \ref{tab1}, the $R^2$ is 0.77. The MSE value is 0.0259, which indicating that the mean error is also low. Figure \ref{fig:dnn_scatter} provides the corresponding scatter plot, showing the correlation between original test values and predictions. Here also, we can see the variance between the two axis components. 

\begin{table}[htbp]
    \centering
    \caption{Metric Evaluation for Different Regression Models \newline
    (Training Size = 60\,\%)}
    \label{tab1}
    
    \begin{tabular}{lccccccc}
    \toprule
        Model & MSE & EVS & $R^2$ \\
    \midrule
        DNN & 0.025995 & 0.770322 & 0.770169 \\
        
        SVR & 0.027359 & 0.690909 & 0.689758 \\
            
    \bottomrule
    \end{tabular}
\end{table}

\subsection{DNN vs SVR}

Metrics form Table \ref{tab1} indicate a dominant performance of DNN in terms of $R^2$, EVS and MSE. The score EVS is used to measure the discrepancy between model-driven values and actual data. The high value near to 1 shows the model is providing a valuable prediction. It appears that the DNN model performs better than SVR. But other facts need to be highlighted. In Table \ref{tab2}, the time needs to execute different models had compared. The fault injection campaign over 1202 flip-flops of the Ethernet-MAC circuit took nearly five days per Modelsim software. SVR seems to be very fast while the DNN needs to optimize a comparatively large set of parameters, as explained in Table \ref{tab3}. But, when compared to traditional fault injection methods, it should be considered that ML/DL models depend 60\% true detests that generated by traditional fault injection methods. 

\begin{table}[htbp]
    \centering
    \caption{Time Performance of Different Models}
    \label{tab2}
    
    \begin{tabular}{lccccccc}
    \toprule
        Model & Time \\
    \midrule
        Fault Injection (1 Modelsim) & 5 days  \\
        Fault Injection (7 Modelsim) & 17 hours \\
        SVR & $<$ 1 minute \\
        DNN & 6 minutes  \\
    \bottomrule
    \end{tabular}
\end{table}

\begin{figure}[ht]
    \includegraphics[clip,trim=50 25 0 0,scale=0.26]{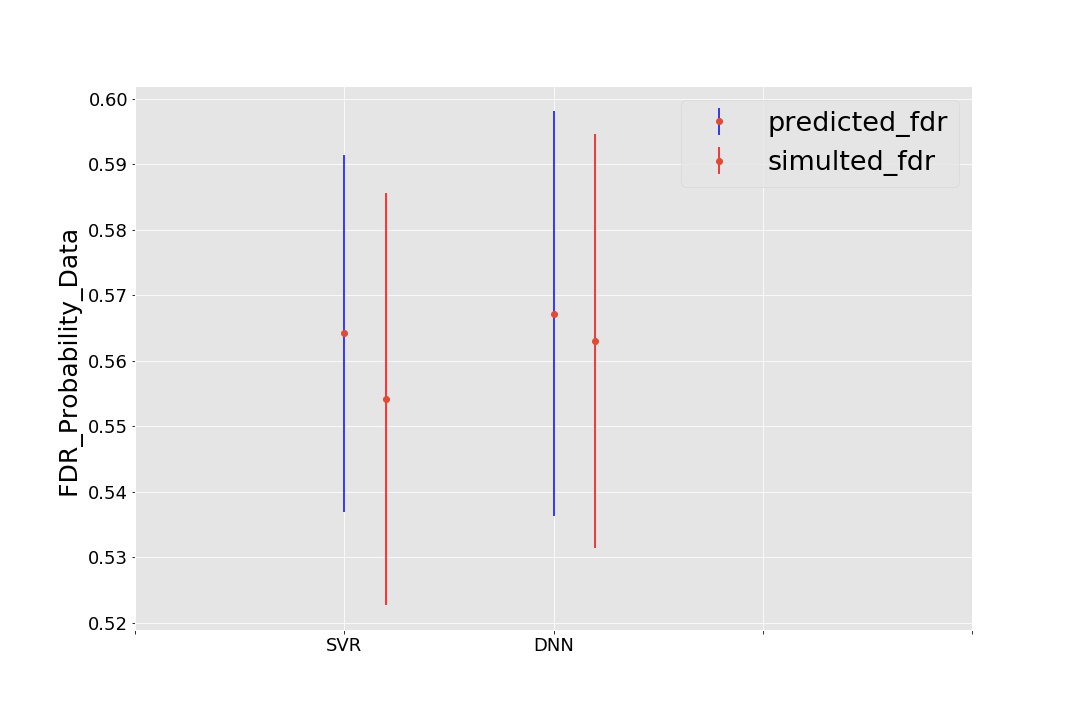}
    \caption{CI comparison : SVR Vs DNN}
    \label{fig:CI}
\end{figure}

Finally, the DNN and SVR have been compared using 95\% Confidence Interval (CI) and Mean values between predicted and original values. This comparison showed in figure \ref{fig:CI}. Here DNN performs comparatively better because the difference between the means of the respective predicted and the target values is small compared to that of SVR.  

\section{Future work}

\begin{figure}[ht!]
  \centering
  \includegraphics[width=\linewidth]{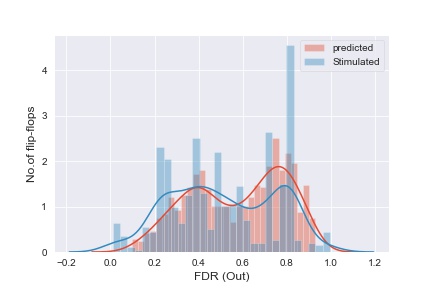}
  \caption{Histogram comparison of GCN model }
    \label{figure_3_1}
\end{figure}

Even though the implemented ML/DL models are achieving their reasonable accuracy within a very short interval of time, all those algorithms need high quality training data. In our case, we have used a $50\% - 60\% $ of the database obtained through first principles methods (fault simulation) for the training process. The real-time data processing applications will not accept this fact to an extent. Every circuit and its electrical characteristics vary from one to another. It could be useful to solve the issue of the training data set by using Graph Convolutional Neural Network (GCN)\cite{kipf2017semi}, that needs only $5\% - 10\%$ of training data. Recent research work about this idea had published in \cite{8792929}. Another development direction is to develop acceptable prediction over gate-level netlist with the node2vec feature matrix and with more advanced graph-based deep neural architectures. A GCN based probability distribution comparison has shown in \ref{figure_3_1}. According to the comparison, a graph convolutional neural network can reach an acceptable prediction goal.
 
\section{Conclusion}

The works implemented in this paper depict the importance of extracting a low-dimensional feature matrix from a gate-level netlist of logic circuits by the node2vec algorithm. This feature matrix has validated using SVR and DNN machine learning algorithms. These algorithms have compared with different regression metrics and diagrams. The whole experiment is proving that the extracted feature matrix by the node2vec algorithm, can be used to perform ML/DL algorithms successfully. This feature space can also apply to complex neural network architectures to reduce the estimation time of different circuit reliability factors.

\bibliographystyle{IEEEtran}
\bibliography{BIBLIOGRAPHY/IEEE_NORCAS_2019}

\end{document}